\documentclass[paper]{ieice}

\usepackage{cite}
\usepackage[dvips]{graphicx}
\usepackage{amsmath}
\usepackage{times}
\usepackage{latexsym}
\usepackage{graphicx}
\usepackage{bm}
\usepackage{amssymb}
\usepackage[center]{caption2}
\usepackage{stfloats}
\usepackage{cases}
\usepackage{array}
\usepackage{setspace}
\usepackage{fancyhdr}

\field{B} \vol{89} \no{4}

\title{Frequency Offset Estimation for OFDM Systems
with a Novel Frequency Domain Training Sequence}

\authorlist{
 \authorentry[yxjiang@seu.edu.cn]{Yanxiang Jiang}{s}{label1}
 \authorentry{Xiqi Gao}{n}{label1}
 \authorentry{Xiaohu You}{n}{label1}
}
\affiliate[label1]{The authors are with the National Mobile
Commun. Research Lab., SouthEast University, NanJing, 210096
China. }

\received{2005}{5}{16}
\revised{2005}{10}{20}

\allowdisplaybreaks[4]
\newtheorem{theorem}{Theorem}

\newcaptionstyle{mystyle}{
  \captionlabel \: \captiontext \par}
\captionstyle{mystyle}

\newcaptionstyle{myanotherstyle}{
\centering  \captionlabel \: \captiontext \par}
\captionstyle{myanotherstyle}

\begin{document}
\maketitle
\begin{summary}
  A novel frequency domain training sequence and the corresponding
carrier frequency offset (CFO) estimator are proposed for
orthogonal frequency division multiplexing (OFDM) systems over
frequency-selective fading channels. The proposed frequency domain
training sequence comprises two types of pilot tones, namely
distinctively spaced pilot tones with high energies and uniformly
spaced ones with low energies. Based on the distinctively spaced
pilot tones, integer CFO estimation is accomplished. After the
subcarriers occupied by the distinctively spaced pilot tones and
their adjacent subcarriers are nulled for the sake of interference
cancellation, fractional CFO estimation is executed according to
the uniformly spaced pilot tones. By exploiting a predefined
lookup table making the best of the structure of the distinctively
spaced pilot tones, computational complexity of the proposed CFO
estimator can be decreased considerably. With the aid of the
uniformly spaced pilot tones generated from Chu sequence with
cyclically orthogonal property, the ability of the proposed
estimator to combat multipath effect is enhanced to a great
extent. Simulation results illustrate the good performance of the
proposed CFO estimator.
\end{summary}


\begin{figure*}[b]
\centering
\includegraphics[scale=0.65]{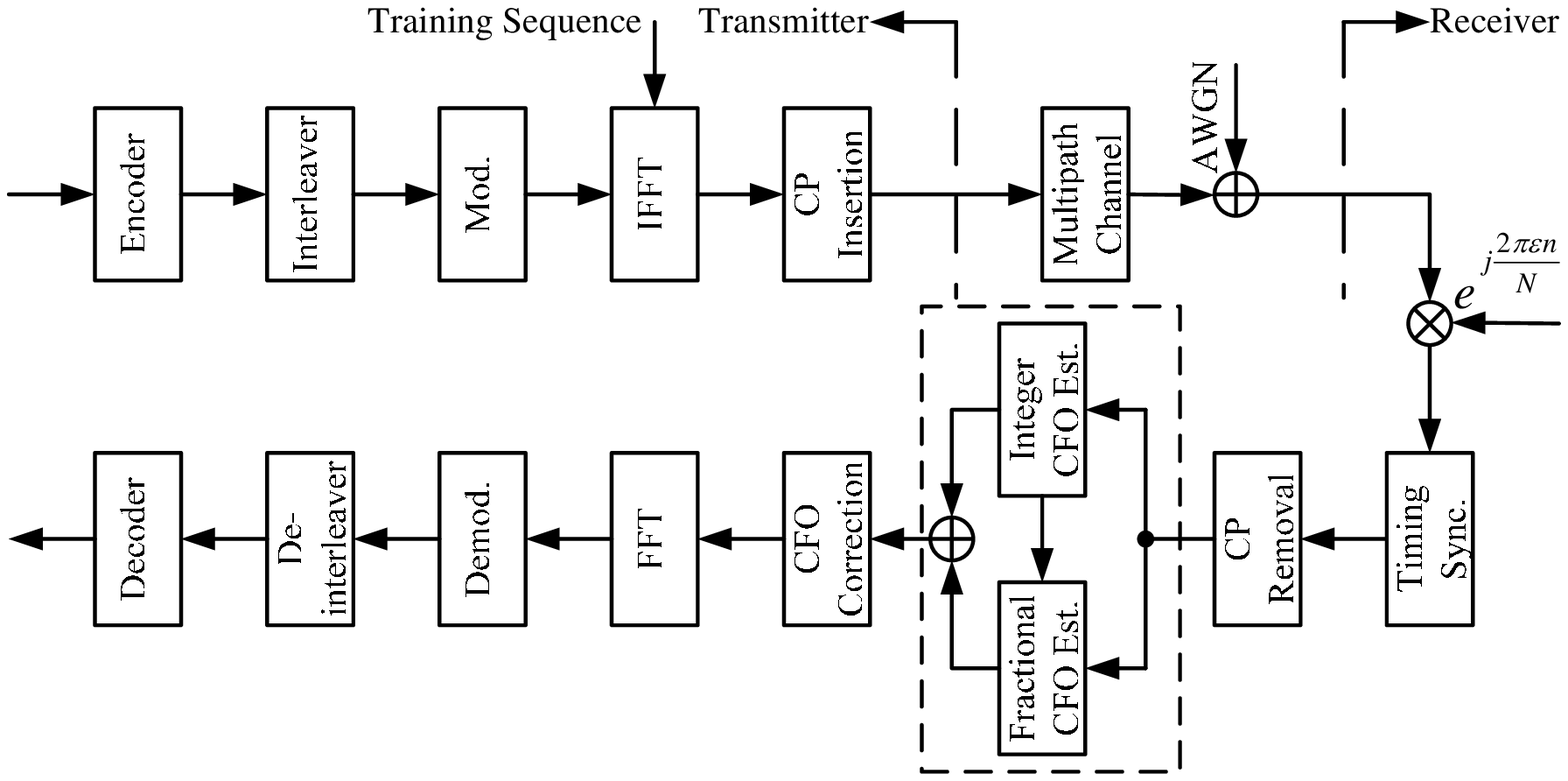}
\captionstyle{myanotherstyle} \caption{Baseband-equivalent OFDM
system model with the proposed CFO estimator.} \label{fig1}
\end{figure*}

\section{Introduction}

    The tendencies of modern wireless mobile communication are
faster transmission rate, better quality of service (QoS), higher
spectrum efficiency and larger system capacity. Orthogonal
frequency division multiplexing (OFDM) has technology predominance
to achieve the above objectives \cite{Bingham}. It has been
adopted successfully in several wireless standards such as digital
audio broadcasting (DAB), digital video broadcasting-terrestrial
(DVB-T), Hiperlan/2, IEEE 802.11a and IEEE 802.16a. It is also a
potential candidate for beyond 3G (B3G) wireless mobile systems.
However, as a parallel transmission technique, OFDM is much more
sensitive to frequency domain interferences such as carrier
frequency offset (CFO) than single carrier (SC) technique
\cite{Pollet}.

    Many excellent works on the topic of CFO estimation have
been reported by other researchers \cite{Beek,Liu,
Tureli1,Moose,Schmidl,Morelli1,Morelli2,Xiaoli,Nogami,Zhang,
Lei,Yanxiang1}. Among these contributions, according to whether
the CFO estimators use training sequences or not, they could be
classified as blind ones \cite{Beek,Liu,Tureli1} and data-aided
ones \cite{Moose,Schmidl,Morelli1,Morelli2,Xiaoli,Nogami,Zhang,
Lei,Yanxiang1}. Focusing on blind approaches, Beek develops a
maximum likelihood (ML) estimator in \cite{Beek} by exploiting the
redundancy in the cyclic prefix (CP). Liu and Tureli take
advantage of the known OFDM subspace structure due to the
placement of virtual subcarriers and propose blind estimation
methods reminiscent of spectral analysis techniques in array
processing, i.e., MUSIC and ESPRIT \cite{Liu,Tureli1}. Opposite to
blind approaches, data-aided CFO estimation can be accomplished
with the aid of training sequences. In \cite{Moose}, an ML CFO
estimator employing repeated training sequences is developed by
Moose. Later, Schmidl and Cox propose a robust frequency
synchronization method based on a training sequence with identical
halves \cite{Schmidl}. The solution of \cite{Schmidl} is
generalized by exploiting a training sequence with multiple
identical parts \cite{Morelli1,Morelli2}. All the above mentioned
data-aided CFO estimators are employing time domain training
sequences with repetitive structure. On the other hand, quite a
bit of research on OFDM is devoted to CFO estimation based on
frequency domain training sequence
\cite{Xiaoli,Nogami,Zhang,Lei,Yanxiang1}. In \cite{Xiaoli}, Ma and
Giannakis propose an adaptive gradient descent method for CFO
estimation based on null subcarriers. Here, for description
convenience, we treat the null subcarriers as a kind of special
`training sequence'.  In \cite{Nogami,Zhang}, Nogami and Zhang
introduce fast Fourier-transform (FFT) based CFO estimation
methods with proper window functions on the relevant
frequency-assignment schemes. Lei proposes a consistent ML CFO
estimator in \cite{Lei}, which exploits the relationship between
CFO and the periodogram of the frequency domain training sequence.
By employing the so-called time-frequency training sequence, a low
complexity CFO estimator is introduced in \cite{Yanxiang1}.

    Concentrating on Lei's estimator in \cite{Lei}, we find that
different from previous $ad \ hoc$ estimators, this kind of CFO
estimator is systematically derived. Analysis demonstrates that
this kind of ML estimator is asymptotically unbiased and
efficient. However, the computational complexity of Lei's
estimator, which exploits large point FFT grid-searching to
improve the estimation performance, is rather high. Furthermore,
since the residual CFO is approximately calculated by exploiting
the magnitude attenuation in the vicinities of those CFO-shifted
pilot tones, the estimate precision can't be guaranteed in radio
channels with large multipath spread and small coherence
bandwidth. In the above channels, the adjacent pilot tones with
frequency separation greater than the coherence bandwidth may
undergo different fading. When the current pilot tone suffers from
deep fading, the adjacent pilot tones may hardly undergo any
fading. Therefore, the right-hand sides of (41) and (42) in
\cite{Lei} should contain the items concerning the adjacent pilot
tones. Otherwise the errors originating from the approximations
(41), (42) can't be ignored even if the averaging operation in
(44) is performed. Accordingly, the fine CFO estimation of Lei's
estimator can't be obtained with sufficient precision. Besides,
since the side information about the residual CFO in (41) and (42)
don't increase proportionably with the oversize ratio, the
performance improvement of Lei's estimator by increasing the
oversize ratio is not obvious.

    To obtain a precise estimate of the true CFO with low
complexity, we propose a novel frequency domain training sequence
and the corresponding CFO estimator. Different from the training
sequence comprising distinctively spaced pilot tones in \cite{Lei}
, the proposed training sequence is composed of distinctively
spaced pilot tones with high energies and uniformly spaced ones
with low energies. Especially, the uniformly spaced pilot tones
are generated from Chu sequence having constant amplitude and zero
auto-correlation (CAZAC) \cite{Chu}. Moreover, we also propose an
approach to reduce the peak-to-average power ratio (PAPR)
concerning the proposed training sequence with low computational
complexity. With the aid of the proposed training sequence, we
then develop the corresponding CFO estimator which overcomes the
drawbacks of Lei's estimator. Utilizing the structure of the
training sequence, we propose to construct a lookup table storing
all kinds of pilot-spacing combinations for the distinctively
spaced pilot tones. Aided by the distinctively spaced pilot tones
and the predefined lookup table, integer CFO estimation can then
be obtained without the need of large point FFT grid searching.
Accordingly, the complexity issue of Lei's estimator is resolved.
After interference cancellation is done, with the aid of the
uniformly spaced pilot tones, fractional CFO estimation based on
the best linear unbiased estimation (BLUE) principle can be
achieved. Since the uniformly spaced pilot tones are generated
from cyclically orthogonal Chu sequence, the fractional CFO
estimator aided by them has strong ability to combat multipath
effect. Correspondingly, the estimate precision issue of Lei's
estimator in channels with large multipath spread is also
resolved. In conclusion, the proposed CFO estimator assisted by
the novel frequency domain training sequence has fairly good
performance with comparatively little computational complexity.

    The rest of the paper is organized as follows. We begin with
the OFDM system model in Section II. The novel frequency domain
training sequence, which consists of distinctively spaced pilot
tones and uniformly spaced ones, is presented in Section III. The
approach concerning PAPR reduction is also introduced in this
section. And then in Section IV, the corresponding CFO estimator,
which is composed of the integer CFO estimator with great
complexity reduction and the fractional CFO estimator with strong
ability to combat multipath effect, is elaborated. Also included
in this section is the complexity analysis. Simulation results are
shown in Section V. Finally, Section VI concludes this paper.

    Notation: Upper (lower) bold-face letters are used for matrices
(column vectors). Superscript $* $, $T $ and $H $ denote
conjugate, transpose and Hermitian transpose, respectively.
$diag\{{\mathbf{\cdot}}\}$ stands for a diagonal matrix with the
elements within the brackets on its diagonal. ${[\mathbf{x}]}_{m}
$ denotes the $m$-th entry of a column vector ${\mathbf{x}} $.
${[\mathbf{A}]}_{m,n} $ denotes the $(m,n)$-th entry of a matrix
${\mathbf{A}} $. $|| \cdot ||^2 $ represents the Euclidean norm
operation. $\lfloor \cdot \rceil$ denotes the nearest integer that
the number within the brackets is rounded to. $(( \cdot ))_N $
denotes the modulus $N $ operation. $E[\cdot]$ denotes the
expectation operation.

\section{System Model}

    The baseband-equivalent OFDM system model with the proposed CFO
estimator is shown in Fig. \ref{fig1}. At transmitter side, the
training sequence symbol is inserted before fixed number of data
symbols to construct a transmitted frame. The data symbols are
generated by passing the information bits through encoder,
interleaver, $M$-ary modulator ($M=2^{M_c}$) and inverse fast
Fourier-transform (IFFT) module. Let $\mathbf {\tilde{p}}_{N} =
[\tilde{p}_0, \tilde{p}_1, \cdots, \tilde{p}_{N-1}]^T $ denote the
frequency domain training sequence. The training sequence symbol
is directly generated with the application of the $N \times N$
normalized IFFT matrix $\mathbf{F}_N$ to $\mathbf{\tilde{p}}_{N}$,
\begin{equation}
\mathbf{p}_{N}  = \mathbf{F}_N \mathbf{\tilde{p}}_{N}
\end{equation}
In order to prevent possible inter-symbol interference (ISI)
between OFDM symbols, a CP with length $N_g$, which is longer than
the multipath delay spread $L$, is inserted before every symbol.
Then the baseband samples are transmitted through a
frequency-selective fading channel with additive white Gaussian
noise (AWGN).

    At receiver side, it is assumed that the received samples are
affected by the normalized CFO $\varepsilon $, which equals the
actual CFO $\Delta F$ divided by the OFDM subcarrier spacing
$\Delta f$. Firstly, timing synchronization is accomplished, then
CPs are removed. Since the proposed CFO estimator is insensitive
to timing synchronization error, ideal timing synchronization is
assumed in this paper. Let $\mathbf{h}_L = [h_0, h_1, \cdots,
h_{L-1}]^T$ denote the normalized finite impulse response of the
multipath channel in discrete-time equivalent form, let
$\mathbf{H}$ denote the $N \times N$ cyclic matrix with first
column $\mathbf{h}_N = [\mathbf{h}_L, \mathbf{0}_{N-L}]^T $, and
define
\begin{equation*}
\mathbf{\Psi} (\varepsilon )=diag\{1,e^{j2\pi \varepsilon /N } ,
\cdots ,e^{j2\pi (N - 1)\varepsilon /N} \},
\end{equation*}
then the received sequence corresponding to the transmitted
training sequence can be written as follows,
\begin{eqnarray}\label{equ2}
  \mathbf{r} &=& e^{j 2\pi \varepsilon N_g /N}
  \mathbf{\Psi(\varepsilon)} \mathbf{H} \mathbf{p}_{N} +
  \mathbf{w} \nonumber \\
  &=& e^{j 2\pi \varepsilon N_g /N} \mathbf{\Psi(\varepsilon)}
  \mathbf{F}_N \mathbf{\tilde{P}}_N \mathbf{\tilde{h}}_N + \mathbf{w}
\end{eqnarray}
where
\begin{equation*}
\mathbf{\tilde{P}}_N = diag\{\tilde{p}_0, \tilde{p}_1, \cdots,
\tilde{p}_{N-1}\}, \mathbf{\tilde{h}}_N = \mathbf{F}_N^H
\mathbf{h}_N,
\end{equation*}
$\mathbf{w} $ stands for zero-mean AWGN with covariance matrix
equal to $\sigma ^2 \mathbf{I}_N$, $\mathbf{I}_N$ denotes the $N
\times N$ identity matrix. Based on $\mathbf{r}$ in (\ref{equ2}),
the proposed integer CFO estimation and fractional CFO estimation
are performed in turn. With the estimated CFO, CFO correction is
applied to the following data symbols. Finally, the information
bits are obtained by passing the corrected data symbols through
FFT module, demodulator, de-interleaver and decoder, respectively.

    From the above description, we can see that the performance of the
CFO estimator depends on the training sequence $\mathbf{\tilde{p}}
_{N}$. By designing the training sequence properly, we can obtain
the corresponding CFO estimator with good performance and low
complexity.

\section{The Novel Frequency Domain Training Sequence}

\subsection{Training Sequence Construction}
\begin{figure}[b]
\centering
\includegraphics[scale=0.55]{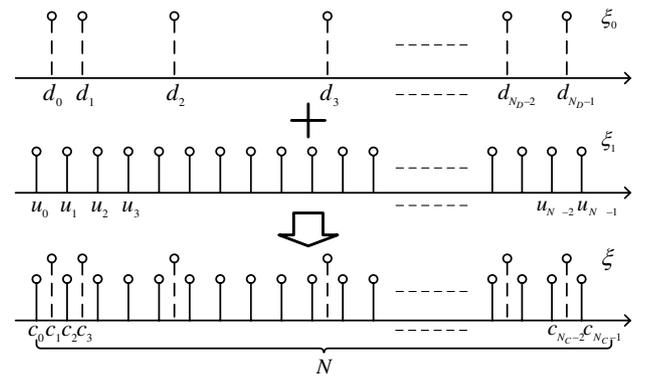}
\captionstyle{mystyle}\caption{Structure of the proposed frequency
domain training sequence with power constraint $\xi=\xi_0 +
\xi_1$. Power $\xi_0$ is allocated to the distinctively spaced
pilot tones, while power $\xi_1$ is allocated to the uniformly
spaced ones. X-coordinate and Y-coordinate represent the indices
and amplitudes of the pilots tones respectively.} \label{fig2}
\end{figure}

    We propose to construct the frequency domain training sequence
$\mathbf{\tilde{p}}_N$ from two types of pilot tones, namely
distinctively spaced ones with high energies and uniformly spaced
ones with low energies. The structure of the proposed training
sequence is illustrated in Fig. \ref{fig2}. The ratio of the total
power of the distinctively spaced pilot tones $\xi_0$ to that of
the uniformly spaced ones $\xi_1$ is set to $\xi_0 :
\xi_1=\alpha:(1-\alpha) $ under constant power constraint $\xi=
\xi_0+ \xi_1 = N $.

    Let ${\mathcal{D}}$ denote the set of the indices for the $N_D$
distinctively spaced pilot tones ${\{\tilde{p}_{d_k}\}_0^{N_D-1}}$
and let $\mathcal{U}$ denote the set of the indices for the $N_U$
uniformly spaced pilot tones ${\{\tilde{p}_{u_k}\}_0^{N_U-1}}$
with $N_U > N_D$. Then, the set of the indices for the $N_C$
non-zero pilot tones ${\{\tilde{p}_{c_k}\}_0^{N_C-1}}$ denoted by
$\mathcal{C}$ can be defined as $\mathcal{C}=\mathcal{D} \cup
\mathcal{U}$, where $\mathcal{D}\cap \mathcal{U}=\phi$. In order
to recover the channel exactly by exploiting the proposed training
sequence, the number of the non-zero pilot tones $N_C$ should be
no less than the multipath delay spread of the channel $L$
\cite{Negi}.

    For the distinctively spaced pilot tones, their indices are designed
to satisfy the following condition,
\begin{multline}\label{dk_condition1}
\hspace*{-20pt}((d_{((n+1))_{N_D}} - d_{n}))_N \ne
((d_{((m+1))_{N_D}} - d_{m}))_N,\\
\textrm{if}\; d_{n} \ne d_{m} ,\; \textrm{and} \;
d_{n} ,d_{m} \in \mathcal{D}
\end{multline}
Suppose $((N))_{2N_U} = 0$, let $X = N/ N_U$ denote the subcarrier
spacing between the adjacent uniformly spaced pilot tones, let
$d_n$ denote the index of the $n$-th distinctively spaced pilot
tone and let $u_{n'}$ denote the index of the left neighboring
uniformly spaced pilot tone to the $n$-th distinctively spaced
pilot tone, then it can be observed from Fig. \ref{fig2} that the
following relationship holds,
\begin{equation}\label{dkindex}
d_n = u_{n'} + \upsilon, \quad \textrm{for} \ \upsilon \in [1,X-1]
\end{equation}
Assume $((N_U))_2 = 0$ and let $P_{ICI,d_k}$ denote the average
power of the total inter-carrier interference (ICI) that
$\tilde{p}_{d_k }$ imposes on the $N_U$ uniformly spaced pilot
tones at receiver side, then we have the following theorem (whose
proof is shown in the appendix),
\begin{theorem}\label{theorem1}
With the proposed training sequence $\mathbf{\tilde{p}} _N$, the
optimum value of $\upsilon$ within the range $[1,X-1]$ which makes
$P_{ICI,d_k}$ achieve its minimum is $X/2$.
\end{theorem}
According to the above theorem, we set $\upsilon = X/2$ in
(\ref{dkindex}) for the proposed training sequence, which helps to
decouple the two types of pilot tones with least interference at
receiver side. Furthermore, since the distinctively spaced pilot
tones with high energies are exploited for the integer CFO
estimation as shown in the following section, their amplitudes are
made constant without loss of generality.

    For the uniformly spaced pilot tones, they are constructed from
a length-$N_U$ Chu sequence $\mathbf{s}_{N_U}$ which has CAZAC.
Let $\mathbf{\tilde{s}}_{N_U} = \mathbf{F}_{N_U}^H
\mathbf{s}_{N_U}$, then the uniformly spaced pilot tones can be
expressed as follows,
\begin{equation}\label{pu}
\tilde{p}_{u_k} = \sqrt{(1-\alpha) X} [\mathbf{\tilde{s}}_{N_U}
]_k, \  \mathrm{for} \; k = 0,1, \cdots, N_U-1.
\end{equation}
Without loss of generality, we set the set of the indices for the
uniformly spaced pilot tones to $\mathcal{U} = \{kX \} _{k=0}
^{N_U-1} $.  An appealing feature has been shown in
\cite{Milewski} that a sequence has CAZAC if and only if its
discrete Fourier-transform (DFT) has CAZAC and vice versa.
Therefore, the uniformly spaced pilot tones generated from Chu
sequence according to (\ref{pu}) also have constant amplitude.

\subsection{Peak-to-Average Power Ratio Reduction}

    Since the PAPR problem often needs to be resolved for low
cost linear power amplifier at transmitter side in many systems
that utilize frequency domain for data recovery and the amplifier
efficiency increases monotonically as the PAPR decreases [18, Fig.
1], the next question considered here is how to make PAPR
concerning the proposed training sequence as small as possible.

    Define
\begin{equation*}
\mathbf{\tilde{p}}_{N_D} = [\tilde{p}_{d_0}, \tilde{p}_{d_1},
\cdots, \tilde{p}_{d_{N_D -1}}]^T,
\end{equation*}
\begin{equation*}
\mathbf{\tilde{p}}_{N_U} = [\tilde{p}_{u_0}, \tilde{p}_{u_1},
\cdots, \tilde{p}_{u_{N_U -1}}]^T,
\end{equation*}
\begin{equation*}
\mathbf{F}_{\beta N \times N_D} = [\mathbf{f}_{d_0}^{\beta N},
\mathbf{f}_{d_1}^{\beta N}, \cdots, \mathbf{f}_{d_{N_D -1}}^{\beta
N}],
\end{equation*}
\begin{equation*}
\mathbf{F}_{\beta N \times N_U} = [\mathbf{f}_{u_0}^{\beta N},
\mathbf{f}_{u_1}^{\beta N}, \cdots, \mathbf{f}_{u_{N_U -1}}^{\beta
N}],
\end{equation*}
where $\mathbf{f}_k^{\beta N}$ denotes the $k$-th column of the
$\beta N \times \beta N$ IFFT matrix $\mathbf{F}_{\beta N}$, then
the $\beta$-times oversampled time domain sequence $\mathbf{p}
_{\beta N}$ corresponding to the proposed training sequence can be
written as follows,
\begin{equation}\label{TS_Time}
\mathbf{p}_{\beta N}  = \sqrt{\beta}\mathbf{F}_{\beta N \times
N_D} \mathbf{\tilde{p}}_{N_D} + \sqrt{\beta}\mathbf{F}_{\beta N
\times N_U} \mathbf{\tilde{p}}_{N_U}
\end{equation}
It is well known that the PAPR of the continuous-time signal can't
be obtained precisely by the use of Nyquist rate sampling, which
corresponds to the case of $\beta = 1$. It is shown in
\cite{Tellambura} that $\beta = 4$ can provide sufficiently
accurate PAPR results. The PAPR computed from the $\beta$-times
oversampled time domain sequence $\mathbf{p}_{\beta N}$ is given
by
\begin{equation}
{PAPR} = \mathop {\max }\limits_{n \in [0,\beta N - 1]} \{
|[\mathbf{p}_{\beta N}]_n |^2 \}
\end{equation}

    After $N_D,N_U,\alpha,\mathcal{U}$ are all decided, we need to design
$\mathcal{D}$ and $\mathbf{\tilde{p}}_{N_D}$ to obtain a small
PAPR concerning the proposed training sequence. Under the
constraints that the elements of $\mathcal{D}$ satisfy the
conditions (\ref{dk_condition1}), (\ref{dkindex}) with $\upsilon =
X/2$ and that the elements of $\mathbf{\tilde{p}}_{N_D}$ have
constant amplitude, the optimal $\mathcal{D}$ and
$\mathbf{\tilde{p}}_{N_D}$ can be obtained by solving the
following $minmax$ problem,
\begin{equation}\label{PAPR1}
\{ \mathcal{D} ,\tilde{ \mathbf{p}}_{N_D } \}  = \mathop {\arg
\min }\limits_{\mathcal{D} ,\tilde{ \mathbf{p}}_{N_D } }
\left\{\mathop {\max }\limits_{n \in [0,\beta N - 1]} \{
|[\mathbf{p}_{\beta N}]_n |^2 \}\right\}
\end{equation}

    In principle, the elements of $\mathcal{D}$ should be distributed
in the range $[0,N-1]$ as `evenly' as possible in order to track
the variety of the fading channel accurately. After $\mathcal{D}$
is determined according to the above principle, the $2N_D$-
dimensional optimization problem concerning PAPR in (\ref{PAPR1})
is reduced to a $N_D$-dimensional optimization one. Furthermore,
since the integer CFO estimation as shown in the following section
is only relevant to the amplitudes of the distinctively spaced
pilot tones, in order to further simply the $N_D$-dimensional
optimization problem, we put the following constraint on the
distinctively spaced pilot tones,
\begin{equation}
\tilde{p}_{d_k} = \sqrt{\frac{\alpha N}{N_D}} (-1)^{i_k},
\mathrm{for} \; i_k \in \{0,1\}, k = 0,1,\cdots, N_D-1.
\end{equation}
Therefore, $\tilde{ \mathbf{p}}_{N_D }$ is determined solely by
$[i_0,i_1,\cdots,i_{N_D-1}]$ with definite $\alpha, N_D$ and $N$.
Correspondingly, we can express the above $N_D$-dimensional
optimization problem concerning PAPR as follows,
\begin{multline}\label{papr2}
\hspace*{-15pt}[i_0,i_1,\cdots,i_{N_D-1}]  = \\
\mathop {\arg \min}\limits_{[i_0,i_1,\cdots,i_{N_D-1}] \in
\{0,1\}^{N_D} } \left\{\mathop {\max }\limits_{n \in [0,\beta N -
1]} \{ |[\mathbf{p}_{\beta N}]_n |^2 \}\right\}
\end{multline}
Compared with the approach as shown in (\ref{PAPR1}), the above
approach has rather less computational complexity. Effectiveness
of the above approach will be shown through simulation results in
Section V.

\section{Low Complexity OFDM CFO Estimator}

    The CFO estimation based on the proposed training sequence is
separated into two phases: the integer and fractional CFO
estimation. In this section, the integer CFO estimator based on
the distinctively spaced pilot tones with great complexity
reduction and the fractional CFO estimator based on the uniformly
spaced ones with strong ability to combat multipath effect will be
described in detail separately. The corresponding complexity
analysis will also be presented subsequently.

\subsection{Integer CFO Estimator Based on the Distinctively
Spaced Pilot Tones}

    Define
\begin{equation*}
\mathbf{F}_{N \times N_C} = [\mathbf{f}_{c_0 }^N ,\mathbf{f}_{c_1
}^N , \cdots ,\mathbf{f}_{c_{N_C - 1}}^N],
\end{equation*}
\begin{equation*}
\mathbf{\tilde{P}}_{N_C} = diag\{\tilde{p}_{c_0}, \tilde{p}_{c_1},
\cdots, \tilde{p}_{c_{N_C-1}}\},
\end{equation*}
\begin{equation*}
\mathbf{\tilde{h}}_{N_C} =
[[\mathbf{\tilde{h}}_N]_{c_0},[\mathbf{\tilde{h}}_N]_{c_1},
\cdots, [\mathbf{\tilde{h}}_N]_{c_{N_C-1}}]^T,
\end{equation*}
then the received sequence corresponding to the proposed training
sequence can be written as follows,
\begin{equation}\label{equ112}
  \mathbf{r} = \mathbf{A} (\varepsilon)\mathbf{\tilde{h}}_{N_C}
  + \mathbf{w}
\end{equation}
where
\begin{equation*}
\mathbf{A} (\varepsilon) = e^{j 2\pi \varepsilon N_g /N}
\mathbf{\Psi(\varepsilon)} \mathbf{F}_{N \times N_C}
\mathbf{\tilde{P}}_{N_C}.
\end{equation*}
With the assumption that $\mathbf{w}$ is zero-mean AWGN with
covariance matrix equal to $\sigma ^2 \mathbf{I}_N$, the
log-likelihood function of $\mathbf{r}$ conditioned on
$\varepsilon$ and $\mathbf{\tilde{h}} _{N_C}$ can be obtained as
(ignoring the constant items),
\begin{equation}\label{likelihoodf}
\mathrm{ln}\; p(\mathbf{r} |\varepsilon ,\mathbf{\tilde{h}}
_{N_C}) =  - \sigma ^{ - 2} || {\mathbf{r} - \mathbf{A}
(\varepsilon)\mathbf{\tilde{h}}_{N_C} }||^2
\end{equation}
Fixing $\varepsilon $ and maximizing (\ref{likelihoodf}) with
respect to $\mathbf{\tilde{h}}_{N_C} $, we can get the ML estimate
of $\mathbf{\tilde{h}}_{N_C} $ \cite{Kay},
\begin{eqnarray}
\mathbf{\hat{h}}_{N_C} &=& [\mathbf{A}^H (\varepsilon) \mathbf{A}
(\varepsilon)]^{-1} \mathbf{A}^H (\varepsilon) \mathbf{r}
\nonumber \\
&=& e^{-j 2\pi \varepsilon N_g /N} \mathbf{\tilde{P}}_{N_C}^{-1}
 \mathbf{F}_{N \times N_C}^H \mathbf{\Psi}^H (\varepsilon)\mathbf{r}
\end{eqnarray}
Substituting $\mathbf{\hat{h}}_{N_C} $ into (\ref{likelihoodf}),
after some straightforward manipulations and dropping the
irrelevant items, we can obtain the reformulated log-likelihood
function conditioned on $\varepsilon$ as follows,
\begin{equation}\label{LikelihoodFunc}
\mathrm{ln}\; p(\mathbf{r} |\varepsilon ) = ||\mathbf{F}_{N \times
N_C}^H \mathbf{\Psi} ^H (\varepsilon)\mathbf{r}||^2
\end{equation}
Therefore, the ML estimate of $\varepsilon$ is given by
\begin{eqnarray}\label{equml}
\hat \varepsilon &=& \mathop {\arg\max }\limits_{\hat{\varepsilon}
} \{ ||\mathbf{F}_{N \times N_C}^H \mathbf{\Psi} ^H (\hat
\varepsilon)\mathbf{r}||^2 \} \nonumber \\
&=& \mathop {\arg\max }\limits_{\hat{\varepsilon} } \{
{\sum\limits_{k = 0}^{N_C - 1} {\Xi ( \hat{\varepsilon}_{k} })}\}
\end{eqnarray}
where
\begin{equation*}
\hat{\varepsilon}_{k} = \hat{\varepsilon} + c_k,
\Xi(\hat{\varepsilon}_{k})=\frac{1}{N}| {\sum\limits_{n = 0}^{N -
1} {[\mathbf{r}]_n e^{ - j2\pi n \hat{\varepsilon}_{k} /N} } }
|^2.
\end{equation*}
We point out here that $\Xi(f)$ denotes the periodogram of the
received sequence $\mathbf{r}$ with period $N$. Suppose that at
least two neighboring distinctively-spaced pilot tones and two
uniformly-spaced pilot tones with indices $u_n$ and $u_m$ do not
coincide with channel nulls of the frequency-selective fading
channels, where $u_n$ and $u_m$ satisfy the following condition,
\begin{equation}
(u_n - u_m) \notin \{d| d = d_k - d_l, \ \textrm{for}\  d_k \neq
d_l \ \mathrm{and}\  d_k, d_l \in \mathcal{D}\}.
\end{equation}
Then exploiting the similar approach as in \cite{Lei}, we can
prove that $\hat \varepsilon = \varepsilon $ is the unique value
to maximize $\mathrm{ln}\; p(\mathbf{r} |\varepsilon )$ for
$\hat{\varepsilon}, \varepsilon \in (-N/2,N/2]$ with the proposed
training sequence $\mathbf{\tilde{p}} _N$. Accordingly, the
estimation range of the CFO estimator in (\ref{equml}) is $(-N/2,
N/2]$.

    Furthermore, it can be seen that $\Xi(f)$ can be computed through
FFT technique. In order to simply the computation, we only invoke
$N$-point FFT over $\mathbf{r}$. Define
\begin{equation}
\tilde{r}_{((k'))_N} = \mathbf{f}_{((k'))_N}^{N,H} \mathbf{r}, \
\mathrm{for} \; k'=-\frac{N}{2}+1, - \frac{N}{2}+2,
\cdots,\frac{N}{2}.
\end{equation}
then we can obtain the integer CFO estimate of $\varepsilon$ as
follows,
\begin{equation}\label{ccfoe}
\hat {\varepsilon}_I =  \mathop {\arg\max }\limits_{k' } \{
{\sum\limits_{k = 0}^{N_C - 1} {|\tilde{r}_{((k'+c_k))_N}|^2} } \}
\end{equation}

    Notice that the computational complexity of the integer CFO
estimation in (\ref{ccfoe}) is still very high. In order to
further simplify the computation, we take the received sequence
transformed from time domain to frequency domain into
consideration. With the assumption that $N $, $X$ and the
signal-to-noise ratio (SNR) are all large enough, the ICI
resulting from adjacent non-zero pilot tones can be ignored and
the following approximation can be achieved,
\begin{multline}\label{Approx}
\hspace*{-17pt}| \tilde{r}_{(( c_k +\lfloor\varepsilon\rceil))_N}
| \doteq |\tilde{p}_{ c_k }| | [\mathbf{\tilde{h}}_N]_{c_k }|
{sinc{(\varepsilon -\lfloor\varepsilon \rceil)}} \hfill\\
\times |1 + \frac{{[\tilde{\mathbf{w}}]_{((c_k +\lfloor\varepsilon
\rceil))_N } }}{\tilde{p}_{ c_k } [\mathbf{\tilde{h}}_N]_{c_k }
sinc{ (\varepsilon -\lfloor\varepsilon \rceil )}}e^{ - j\frac{\pi
(\varepsilon -\lfloor\varepsilon \rceil ) (N - 1) +2\pi
\varepsilon N_g}{N}}| \hfill\\
\doteq  | {\tilde{p}_{ c_k } } || [\mathbf{\tilde{h}}_N]_{c_k }|
{sinc{(\varepsilon -\lfloor\varepsilon \rceil)}} \hfill
\end{multline}
where
\begin{equation*}
\tilde{\mathbf{w}} = \mathbf{F}_N^H \mathbf{w}, sinc(x) = sin(\pi
x)/(\pi x).
\end{equation*}
Taking the above approximation into consideration, we can get the
suboptimal integer CFO estimation as follows,
\begin{multline}\label{sopccfoe}
\hspace*{-17pt}\hat {\varepsilon}_I  = \\
\mathop {\arg\max }\limits_{k'} \{ {\sum\limits_{k = 0}^{N_D - 1}
{|\tilde{r}_{((k'+d_k))_N}|^2} +\sum\limits_{k = 0}^{N_U - 1}
{|\tilde{r}_{((k'+u_k))_N}|^2}}\}  \hfill\\
\doteq \mathop {\arg\max }\limits_{k' } \{ {\sum\limits_{k =
0}^{N_D - 1} {|\tilde{r}_{((k'+ d_k))_N}|^2} } \} \hfill
\end{multline}

\begin{table*}[t]
\centering
\begin{minipage}{\linewidth}
\renewcommand{\thempfootnote}{\arabic{footnote}}
\setcounter{footnote}{1} \caption{{Contents of the Predefined
Lookup Table} \protect\footnote{Each row stores a possible
pilot-spacing combination.}} \label{table1}
\begin{center}
\begin{tabular}{c|c|c|c}
\hline $\  d_1  - d_0 \ $
 & $d_2  - d_0 $ & $ \qquad\qquad\cdot  \cdot  \cdot \qquad\qquad$
& $d_{N_D-1}  - d_0 $ \\
\hline $d_2  - d_1 $ & $d_3  - d_1 $ & $ \cdot  \cdot  \cdot $
& $d_0  - d_1 + N $ \\
\hline $ \cdots $ & $ \cdots $ & $ \cdots $ & $ \cdots $ \\
\hline $\ d_0  - d_{N_D-1} + N \ $ & \ $d_1  - d_{N_D-1} + N \ $ &
$ \cdot \cdot \cdot $ & \ $d_{N_D-2}  - d_{N_D-1} + N\  $ \\
\hline
\end{tabular}
\end{center}
\end{minipage}
\end{table*}

    Let
\begin{equation*}
d_{k_0 }  = \mathop {\arg \max }\limits_{d_k  \in \mathcal{D}} \{
|\tilde r_{((d_k  + \lfloor\varepsilon\rceil ))_N} |\},
\end{equation*}
\begin{equation*}
u_{k_1 } = \mathop {\arg \max }\limits_{u_k  \in \mathcal{U}} \{
|\tilde r_{((u_k  + \lfloor\varepsilon\rceil))_N } |\}.
\end{equation*}
Taking (\ref{Approx}) into consideration again, if and only if the
following condition is satisfied,
\begin{equation}
|[\mathbf{\tilde{h}}_N]_{d_{k_0} } /
[\mathbf{\tilde{h}}_N]_{u_{k_1}} |
> \sqrt{(1-\alpha) N_D / (\alpha N_U )} ,
\end{equation}
we can obtain
\begin{equation}\label{condition}
\mathop {\arg \max }\limits_{c_k  \in \mathcal{C}} \{ |\tilde
r_{((c_k  + \lfloor\varepsilon\rceil))_N } |\} \in \mathcal{D}
\end{equation}
Let
\begin{equation}\label{pc}
P_{correct} = P(\mathop {\arg \max }\limits_{c_k \in \mathcal{C}}
\{ |\tilde r_{((c_k  + \lfloor\varepsilon\rceil))_N } |\} \in
\mathcal{D})
\end{equation}
denote the probability that (\ref{condition}) holds. Using
extensive simulations in the next section, we find that by
increasing the value of $N_D$ with $\alpha N_U  / [(1-\alpha)
N_D]$ kept invariable, the probability $P_{correct}$ approaches 1.

\begin{figure}[b]
\centering
\includegraphics[height=0.28\textheight]{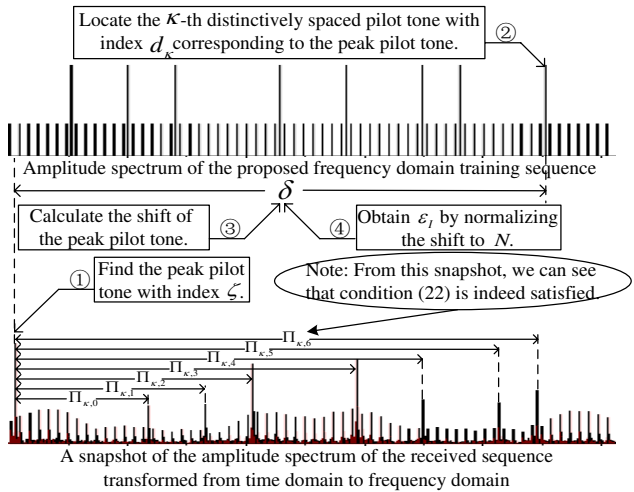}
\captionstyle{mystyle} \caption{Schematic diagram of the proposed
integer CFO estimator with equations (24)-(27).} \label{fig3} 
\end{figure}
    With condition (\ref{condition}), we propose a low complexity
integer CFO estimator as illustrated in Fig.~\ref{fig3}. Firstly,
the peak pilot tone which is affected by the fading channel least
is found,
\begin{equation}
\zeta = \mathop {\arg \max }\limits_{k \in [0,N - 1]} \{|\tilde
r_k | \} \label{CE2}
\end{equation}
Since $((\zeta-\lfloor\varepsilon\rceil))_N$ is a member of the
set $\mathcal{D}$ and the spacings between the adjacent members of
$\mathcal{D}$ are distinctive, we then exploit a lookup table with
size $N_D \times (N_D-1)$, which stores every possible
pilot-spacing combination for the distinctively spaced pilot tones
as shown in Table 1, to locate the distinctively spaced pilot tone
corresponding to the peak pilot tone in $\mathcal{D} $. The
corresponding operation is,
\begin{equation}
\kappa = \mathop {\arg \max }\limits_{k \in [0,N_D - 1]}
\{\sum\limits_{g = 0}^{N_D - 2} |\tilde {r}_{ ((\Pi _{k,g}  +
\zeta ))_{N}} |^2 \} \label{CE3}
\end{equation}
where $\Pi_{k,g} $ denotes the content stored in the $k $-th row
$g $-th column of the lookup table. Then, the shift of the
searched peak pilot tone is calculated,
\begin{equation}
\delta  = \zeta  - d_\kappa  \label{CE4}
\end{equation}
Finally, the integer CFO estimate of $\varepsilon$ can be readily
obtained by normalizing the shift to $N $,
\begin{equation}
\hat{\varepsilon} _I  = \left\{ {\begin{array}{ll}
  { - N + \delta ,} & {\ \mathrm{if} \; \delta  > N/2} , \\
  {N + \delta ,} & {\ \mathrm{else \; if} \; \delta  \leqslant  - N/2},  \\
  {\delta ,} & {\ \textrm{else}} . \\
  \end{array} } \right.\label{CE5}
\end{equation}

\begin{table*}[t]
\centering
\begin{minipage}{\linewidth}
\renewcommand{\thempfootnote}{\arabic{footnote}}
\setcounter{footnote}{1} \caption{ {Computational complexity of
FBE and LNE}} \label{table2}
\begin{center}
\begin{tabular}{c|c|c}
\hline
&\   Number of Real Additions\  & \  Number of Real Multiplications\ \\
\hline \qquad\quad FBE \qquad\quad\ \  & $
\begin{gathered} 4N\log _2 N +N(1.5X+1) +N_D(N_D+1) \end{gathered}
$ &
$\begin{gathered} 4N\log _2 N + N(1.5X+4) +N_D(N_D-2)\end{gathered}$\\
\hline  LNE \protect\footnote{\hspace*{-3pt} For LNE, $L'$ denotes
the oversize ratio of the corresponding FFT interpolation,
\normalsize $M'$ denotes the number of the interpolated signals in
frequency domain whose amplitudes are larger than the predefined
threshold.} & $ 2L'N\log _2 N + 2N_D + N_DM' $ &
$2L'N\log _2 N + 6N_D + N_DM' $\\
\hline
\end{tabular}
\end{center}
\end{minipage}
\end{table*}

\subsection{Fractional CFO Estimator Based on the Uniformly Spaced Pilot Tones}

    After the integer CFO estimation is accomplished, the
distinctively spaced pilot tones and the neighboring ones are
nulled for the sake of cancelling the interference they impose on
the uniformly spaced pilot tones. Let $\tilde{\mathbf{r}}^{ic}$
denote the interference-cancelled frequency domain sequence, then
we can express $\tilde{\mathbf{r}}^{ic}$ according to the above
description as follows,
\begin{equation}\label{null}
[\tilde{\mathbf{r}}^{ic}]_k =
\begin{gathered}
\left\{ {\begin{array}{ll}
{ 0,} & {{\text{ if }}k=((d_{\breve{k}} + \hat{\varepsilon}_I))_N
\; \mathrm{for} \; d_{\breve{k}} \in \mathcal{D}},  \\
{ } & \ { {\text{or}}\ k=((d_{\breve{k}} + \hat{\varepsilon}_I
-1))_N \; \mathrm{for} \; d_{\breve{k}} \in \mathcal{D}} \\
{ } & \ {\hspace*{33pt} \textrm{and} \ \frac
{\sum\nolimits_{d_{\breve{k}} \in \mathcal{D}} {| {\tilde
r_{((d_{\breve{k}} + \hat\varepsilon _I + 1))_N }}|^2 } }
{\sum\nolimits_{d_{\breve{k}} \in \mathcal{D}}{|{\tilde
r_{((d_{\breve{k}} + \hat\varepsilon _I  - 1))_N }}
|^2}} <1 }, \\
{} & \ {{\text{or }} k=((d_{\breve{k}} + \hat{\varepsilon}_I
+1))_N \; \mathrm{for} \; d_{\breve{k}} \in \mathcal{D}} \\
{ } & \ {\hspace*{33pt} \textrm{and} \ \frac
{\sum\nolimits_{d_{\breve{k}} \in \mathcal{D}} {| {\tilde
r_{((d_{\breve{k}} + \hat\varepsilon _I + 1))_N }}|^2 } }
{\sum\nolimits_{d_{\breve{k}} \in \mathcal{D}}{|{\tilde
r_{((d_{\breve{k}} + \hat\varepsilon _I  -
1))_N }}|^2}} >1},\\
{\tilde{r}_k,} & {\ \mathrm{else}.}
\end{array} } \right.
\end{gathered}
\end{equation}
After that, $\tilde{\mathbf{r}}^{ic}$ is transformed from
frequency domain to time domain as follows,
\begin{equation}
\mathbf{r} ^{ic}  = \mathbf{F}_N \tilde{\mathbf{r}}^{ic}
\end{equation}
Then integer CFO correction is carried out on $\mathbf{r} ^{ic}$,
\begin{equation}
\mathbf{r}^{cc}=  e^ {-j \frac{2\pi \hat{\varepsilon}_I N_g} {N}}
\mathbf{\Psi}(-\hat{\varepsilon}_I) \mathbf{r}^{ic}
\end{equation}

    Taking the property of periodical repetition for
$\mathbf{r}^{cc} $ into account, we have
\begin{eqnarray}\label{equ30}
\hspace*{-20pt}R_m \!& = &\! \frac{1}{N - mN_U}\sum\limits_{n =
mN_U}^{N -
1}{[\mathbf{r}^{cc}]_n [\mathbf{r}^{cc}]_{n - mN_U}^* } \nonumber \\
\!& = &\! \sigma_r^2 e^{j2\pi m N_U\varepsilon_F /N}
(1+\varrho_m), \  \textrm{for} \  m \in [0, X/2]
\end{eqnarray}
where
\begin{equation*}
\varepsilon_F = \varepsilon - \hat{\varepsilon}_I,
\end{equation*}
\begin{eqnarray*}
\hspace*{-20pt}\sigma_r^2 &\doteq& \frac{1}{N - mN_U}
\sum\limits_{n = mN_U }^{N - 1} \left\{\{ \sum\limits_{l = 0}^{L -
1} {\{
[\mathbf{s}_{N_U}]_{((n - l ))_{N_U } } h_l } \}\}\right. \\
& & \times \left. \{ \sum\limits_{l = 0}^{L - 1} {\{
[\mathbf{s}_{N_U}]_{((n - mN_U  - l ))_{N_U } } h_l \} } \} ^*
\right\},
\end{eqnarray*}
\begin{multline*}
\hspace*{-20pt}\varrho _m \doteq \frac{1} {\sigma _r^2 (N - mN_U
)}\left\{ \{ \sum\limits_{l = 0}^{L - 1} \{
[\mathbf{s}_{N_U}]_{((n - l ))_{N_U } } h_l
\}\} \breve{w}_{n - m N_U }^{*} \right. \hfill  \\
 \left.+ \{ \sum\limits_{l = 0}^{L - 1} \{
[\mathbf{s}_{N_U}]_{((n - m N_U  - l ))_{N_U } } h_l \}  \} ^*
\breve{w}_n  + \breve{w}_n \breve{w}_{n - m N_U }^{*} \right\}  ,
\end{multline*}
$\breve{w}_n$ denotes zero mean AWGN. With the assumption that the
multipath channel is time-invariant over each OFDM symbol, the
following result can be obtained by exploiting cyclically
orthogonal property of Chu sequence,
\begin{multline}\label{equ31}
\hspace*{-15pt} \sigma_r^2 \doteq \frac{1}{N - mN_U}
\sum\limits_{n = mN_U }^{N - 1} \left\{ \sum\limits_{l = 0}^{L -
1} {\{ |[\mathbf{s}_{N_U}]_ {((n - l ))_{N_U } }|^2 | h_l|^2 }
\}\right. \\
 + \left. \sum\limits_{\scriptstyle l,l' = 0, l \ne l' }^{L - 1}
{\{ [\mathbf{s}_{N_U}]_{((n - l ))_{N_U } }
[\mathbf{s}_{N_U}]_{((n - {l'} ))_{N_U } }^* h_l }
h_{l'}^* \} \right\} \\
\!= \! \frac{1}{N - mN_U} \sum\limits_{n = mN_U }^{N - 1}
\sum\limits_{l = 0}^{L - 1} \left\{{ |[\mathbf{s}_{N_U}]_{((n - l
))_{N_U } }|^2 | h_l|^2 } \right\}
\end{multline}
We can see from (\ref{equ30}) and (\ref{equ31}) that Chu sequence
with cyclically orthogonal property can enhance the ability of the
estimator to combat multipath effect. Actually, with the concept
of zero-correlation zone (ZCZ) in \cite{Fan}, the uniformly spaced
pilot tones generated from a sequence with ZCZ width greater than
$L$, i.e., the multipath delay spread of the channel, is enough.

    Assume that the SNR is sufficiently high and that $|\varepsilon_F|
\leq X/2 $, then the following approximation holds,
\begin{eqnarray}
 \varphi _m & =&  angle( R _m R _{m - 1}^* ) \nonumber\\
& \doteq &  {Imag}(\varrho_m) -Imag(\varrho_{m-1})\nonumber\\
& & \hspace*{-12pt}  + 2\pi N_U\varepsilon_F /N , \quad
\textrm{for} \ m \in [1, X/2]
\end{eqnarray}
where $Imag(\varrho_m)$ denotes the imaginary component of
$\varrho_m$. Accordingly, based on the best linear unbiased
estimation (BLUE) principle \cite{Morelli1,Kay}, fractional CFO
estimation can be obtained as follows,
\setlength\arraycolsep{0.5pt}
\begin{equation}\label{mm}
\hat {\varepsilon} _F  = \frac{N}{2\pi N_U}\sum\nolimits_{m =
1}^{X/2} {\lambda _m \varphi _m }
\end{equation}
where
\begin{equation*}
\lambda _m \!=\! \frac{6(X - m)(X - m + 1) - 0.25X^2}{X(X^2-1)},
\textrm{for} \  m \in [1, X/2].
\end{equation*}

    After the integer and fractional CFO estimation are both accomplished,
the whole CFO is readily obtained as $\hat {\varepsilon}  =
\hat{\varepsilon} _I + \hat{\varepsilon} _F $.

\begin{figure}[b]
\centering
\includegraphics[scale=0.82]{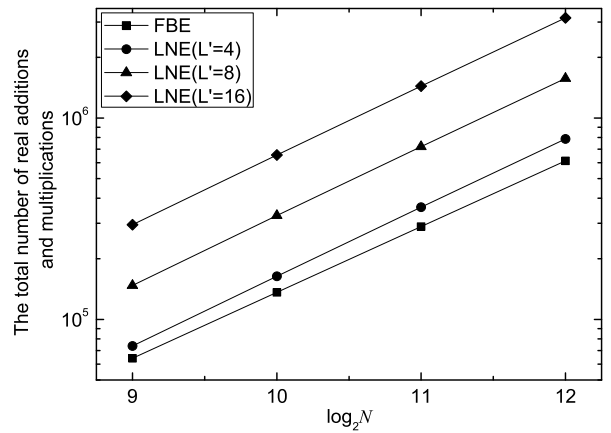}
\captionstyle{mystyle} \caption{Complexity comparison between FBE
and LNE with $L' =4,8,16$.} \label{fig4} 
\end{figure}

\subsection{Complexity Analysis}

    For description convenience, we refer to the proposed estimator
as frequency domain training sequence based estimator (FBE). In
Table 2, the computational complexity of FBE is evaluated in
comparison with that of Lei's estimator (LNE) \cite{Lei}. We
assume that LNE employs a length-$N$ frequency domain training
sequence comprising $N_D$ pilots tones. The comparison is
visualized in Fig. \ref{fig4}, where the total numbers of real
additions and multiplications involved in the two types of
estimators are illustrated as a function of the subcarrier number
$N$. For the considered subcarrier numbers, we set the number of
the distinctively spaced pilot tones to $N_D=N/128$, which
guarantees accurate running of the integer CFO estimator, and set
the subcarrier spacing between the adjacent uniformly spaced pilot
tones to $X=16$, which guarantees little ICI between the two types
of pilot tones in the proposed training sequence at receiver side
and good complexity-performance tradeoff of the fractional CFO
estimator. It can be observed from Fig. \ref{fig4} that the
complexity of FBE is obviously lower than that of LNE.
Furthermore, since the time domain sequence corresponding to the
uniformly spaced pilot tones can be separated into two identical
halves for even $X$, the correlation-based fractional CFO
estimator in \cite{Schmidl}, whose complexity is lower than that
of the fractional CFO estimator in Section IV.B, can be exploited
to further reduce the complexity of FBE but with a slightly worse
performance.

    As far as resource consumption is concerned, we point
out here that the IFFT operation included in the above fractional
CFO estimation can be implemented through FFT technique by making
a conjugate operation on its input and output respectively.
Moreover, since $N$-point FFT operation is also a prerequisite for
an OFDM receiver, the resource that FBE consumes can thus be
reduced considerably.

\section{Simulation Results}

    This section presents the results of Monte Carlo simulations
which illustrate the performance of FBE. Throughout the
simulations, we employ a turbo encoder with code rate $R_c = 1/2$,
an S-random interleaver \cite{Dolinar} with size $20480 $ and a
16-QAM modulator for the OFDM system. Other parameters are set as
follows: carrier frequency $f_c = 5$GHz, bandwidth $B = 10$MHz,
subcarrier number $N=1024$, CP length $ N_g = 64 $, sampling
interval $T_s = 0.1 \mu s$, and symbol duration $T = 108.8 \mu s$.
Two types of multipath Rayleigh fading channels are employed for
the simulations: a 4-path slow fading channel with small delay
spread (Channel 1), a 6-path fast fading channel with large delay
spread (Channel 2). Channel 1 has a classical Doppler spectrum
with maximum Doppler shift $f_d = 50$Hz, while Channel 2 has a
classical Doppler spectrum with $f_d = 200$Hz. For Channel 1, the
relative average-powers and propagation delays of the four paths
are $\{ 0, -9.7,-19.2,-22.8 \}$ dB and $\{0, 0.2, 0.4, 0.8 \} $
$\mu s$, respectively. While for Channel 2, all the six paths have
identical average-powers and their relative propagation delays are
$\{0, 0.3, 0.7, 1.1, 1.3, 2.4 \}$ $\mu s$. For the training
sequence, we design $N_D=8$, $N_U=64$ and $\mathcal{D} = \{ 104,
200, 280, 456, 568, 696, 760, 904 \}$. Besides, according to
(\ref{equ112}), the SNR of the received sequence is defined as
$E_s/N_0 = E[|\mathbf{A}(\varepsilon)
\tilde{\mathbf{h}}_{N_C}|^2]/\sigma ^2$.

\begin{figure}[b]
\centering
\includegraphics[scale=0.85]{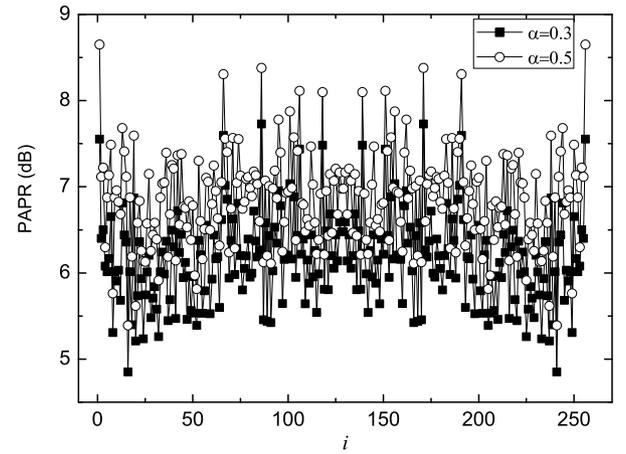}
\captionstyle{mystyle} \caption{PAPR versus $i$ for the proposed
training sequence with $\alpha = 0.3,0.5$.} \label{fig5}
\end{figure}

    With the definition
\begin{equation}
i = \sum\nolimits_{k=0}^{N_D -1} {i_k 2^{N_D -1 -k}},
\end{equation}
Fig. \ref{fig5} presents simulation results of the PAPR
corresponding to each $i$ for the proposed training sequence with
$\alpha = 0.3,0.5$. It can be observed from the figure that the
values of PAPR vary with $i$ in the similar law with different
$\alpha$ and that the PAPR increases with $\alpha$ at the same $i$
for almost all cases. When $\alpha = 0.3$, the minimum PAPR is
achieved at $i= 16$, which correspond to $[i_0, i_1, \cdots,
i_{N_D-1}] = [0 0 0 1 0 0 0 0]$. When $\alpha = 0.5$, the minimum
PAPR is achieved at $i= 241$, which corresponds to $[i_0, i_1,
\cdots, i_{N_D-1}] = [1 1 1 1 0 0 0 1]$. We can also observe from
the figure that 2.88dB and 3.26dB reductions are obtained when the
minimum PAPR is compared with the maximum PAPR for different
$\alpha$.

\begin{figure}[b]
\centering
\includegraphics[scale=0.85]{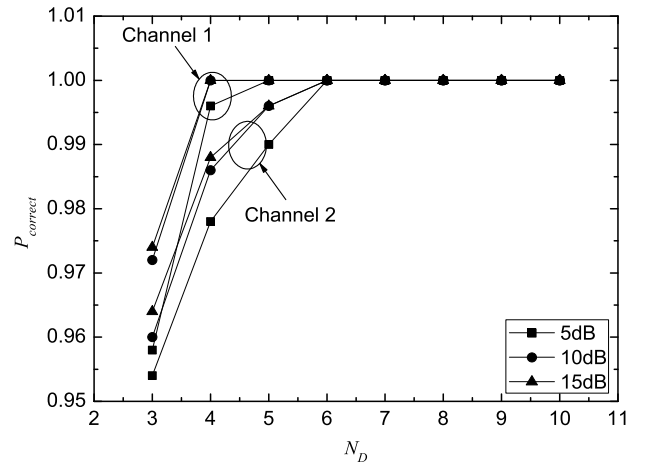}
\captionstyle{mystyle} \caption{The probability $P_{correct}$
versus $N_D$ for the proposed training sequence with $\alpha N_U /
[(1-\alpha) N_D] =24/7, 8$, $E_s/N_0 = 5,10,15$dB in Channel 1 and
Channel 2, respectively.} \label{fig6}
\end{figure}

    Fig.~\ref{fig6} shows the probability $P_{correct} $ defined in
(\ref{pc}) versus $N_D$ for the proposed training sequence with
$\alpha N_U  / [(1-\alpha) N_D] =24/7, 8$ in Channel 1 and Channel
2, respectively. We can see from the figure that by increasing the
value of $N_D$ with $\alpha N_U  / [(1-\alpha) N_D]$ kept
invariable, the probability $P_{correct}$ can always approaches 1.
This verifies the correctness of condition (\ref{condition}).

\begin{figure}[t]
\centering
\includegraphics[scale=0.8]{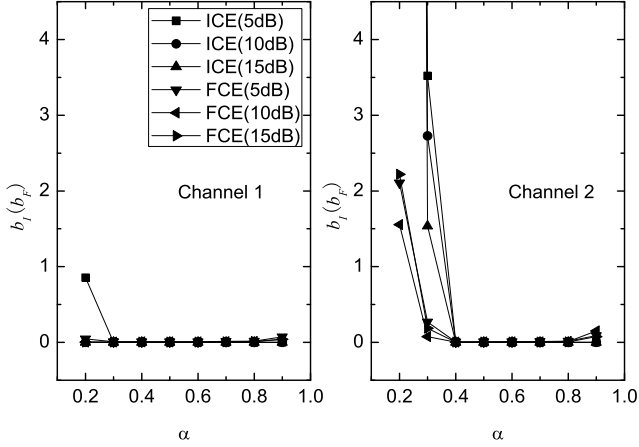}
\captionstyle{mystyle} \caption{Average bias versus $\alpha$ for
ICE and FCE with $E_s/N_0$=5,10,15dB in Channel 1 and Channel 2,
respectively.} \label{fig7}
\end{figure}

    Define the average bias as follows,
\begin{equation}
\begin{gathered}
b_I = \frac{1}{{N_S }}\sum\limits_{k = 0}^{N_S  - 1}
{|(\hat{\varepsilon}_I )_k  - \lfloor\varepsilon\rceil |},\\
b_F = \frac{1}{{N_S }}\sum\limits_{k = 0}^{N_S  - 1}
{|(\hat{\varepsilon}_F )_k  - (\varepsilon -
\lfloor\varepsilon\rceil) |},
\end{gathered}
\end{equation}
where $N_S$ denotes the total number of simulations and
$(\hat{\varepsilon}_I )_k$, $(\hat{\varepsilon}_F )_k$ denote the
estimated integer and fractional CFOs for the $k$-th run,
respectively. For the properly-operating CFO estimator, $b_I$ and
$b_F$ should be around zero. In order to show how the ratio
$\alpha$ impacts the estimate performance, we present in Fig.~
\ref{fig7} the simulation results of the average bias versus
$\alpha$ for the proposed integer CFO estimator (ICE) and
fractional CFO estimator (FCE) with $E_s/N_0$=5,10,15dB in Channel
1 and Channel 2, respectively. It can be seen that ICE works
reliably with large $\alpha$ and that FCE works reliably with
small $\alpha$ under the condition that ICE runs accurately. There
is a safe zone for both of them to operate well. The safe zone is
$[0.3, 0.8]$ in Channel 1, while it is $[0.4, 0.8]$ in Channel 2.

\begin{figure}[t]
\centering
\includegraphics[scale=0.8]{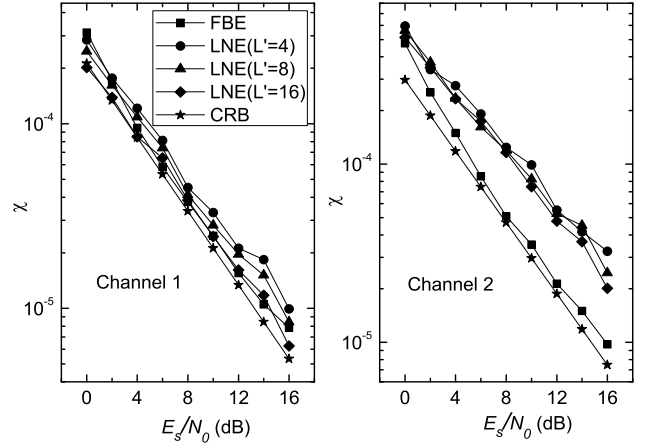}
\captionstyle{mystyle} \caption{MSE versus $E_s/N_0$ for FBE and
LNE with $\alpha=0.3, \varepsilon=9.279$ in Channel 1 and
$\alpha=0.5, \varepsilon=-8.835$ in Channel 2. Also included for
comparison is the CRB.} \label{fig8}
\end{figure}

    To evaluate the performance of FBE, we introduce a variable:
the mean square error (MSE) $\chi$, which is defined as follows,
\begin{equation}
\chi = \frac{1}{N_S}\sum\limits_{k = 0}^{N_S  - 1}
{[(\hat{\varepsilon})_k - \varepsilon]^2}
\end{equation}
where $(\hat{\varepsilon})_k$ denote the estimated CFO for the
$k$-th run. Fig. \ref{fig8} illustrates the MSE performance of FBE
with $\alpha=0.3, \varepsilon=9.279$ in Channel 1 and $\alpha=0.5,
\varepsilon=-8.835$ in Channel 2, respectively. As a benchmark,
the performances of LNE with $L'=4, 8, 16$, where $L'$ denotes the
oversize ratio of the corresponding FFT interpolation, are also
shown in the figure. For LNE, we assume that a frequency domain
training sequence with length $N=1024$ consisting of $N_D=8$ pilot
tones is employed. Also included for comparison is the Cramer-Rao
bound (CRB) \cite{Morelli1} defined as follows,
\begin{equation}\label{CRB}
\text{CRB} = \frac{1.5} {\pi ^2 N(1 - N^{ - 2} )( 1- \alpha )
10^{{\text{SNR}}/10} }
\end{equation}
It can be seen from the figure that the estimate performance of
FBE is similar to that of LNE in Channel 1 and superior to that of
LNE in Channel 2. Thanks to the cyclically orthogonal property of
Chu sequence, FBE is comparatively robust against multipath
channels and its MSE is quite close to the CRB for both of the
considered channels. Whereas the MSE performance of LNE
deteriorates in Channel 2 and there exists a large performance gap
between the MSE of LNE and that of FBE. Recall the complexity
comparison in last section, we can find certain advantages of FBE.

\begin{figure}[t]
\centering
\includegraphics[scale=0.78]{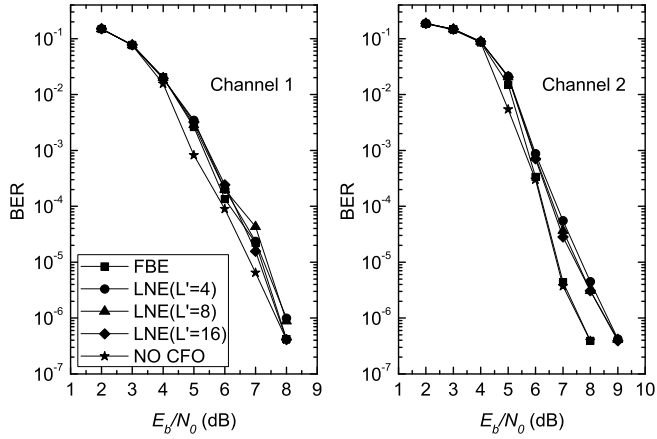}
\captionstyle{mystyle} \caption{BER versus $E_b/N_0$ for FBE and
LNE with $\alpha=0.3, \varepsilon=9.279$ in Channel 1 and
$\alpha=0.5, \varepsilon=-8.835$ in Channel 2. Also included for
comparison is the ideal case when no CFO exists.} \label{fig9}
\end{figure}

    In Fig. \ref{fig9}, we plot the bit error rate (BER)
curves of FBE and LNE as a function of $E_b/N_0$ with $E_b/N_0 =
\frac{E_s}{R_c M_c N_0}$. Also included for comparison is the
ideal case when no CFO exists. For these simulations, we set the
number of inner iterations in the turbo decoder to $6$, and we
also assume ideal channel estimation at receiver side. It can be
observed from the figure that the BER performance of FBE is
similar to that of LNE in Channel 1 and better than that of LNE in
Channel 2, which should be owed to the better MSE performance of
FBE. Therefore, with relatively low complexity and good
performance, FBE is more suitable for practical OFDM systems in
comparison with LNE.

\section{Conclusions}

    In this paper, a novel frequency domain training sequence
composed of distinctively and uniformly spaced pilot tones with
different energies and the corresponding CFO estimator have been
proposed for OFDM systems over frequency-selective fading
channels. The integer CFO estimation is achieved based on the
distinctively spaced pilot tones, while the fractional CFO
estimation is accomplished based on the uniformly spaced ones. The
computational complexity of the proposed CFO estimator has been
decreased considerably by exploiting a predefined lookup table
making the best of the distinctively spaced pilot tones. The
ability of the proposed CFO estimator to combat multipath effect
has also been enhanced greatly with the aid of the uniformly
spaced pilot tones generated from Chu sequence with cyclically
orthogonal property. Moreover, the good performance of the
proposed CFO estimator has been verified through simulations.
Although the CFO estimation discussed here is for OFDM systems,
the proposed estimator is also well-suited for other block
transmission systems such as SC systems.

\appendix*
\section{Proof of Theorem \ref{theorem1}}

    Since integer frequency offset won't introduce ICI and it only has
the effect of shifting the pilot tones, we assume $\varepsilon$ to
be in the range $[-0.5, 0.5]$ in this appendix.

    With the $N \times N$ FFT matrix $\mathbf{F}_N^H$ applied to $\mathbf{r}$
in (\ref{equ2}), we have
\begin{equation}
\tilde{\mathbf{r}} = \mathbf{F}_N^H \mathbf{r}
\end{equation}
Exploiting the proposed frequency domain training sequence
$\mathbf{\tilde{p}} _N$, we can express the non-zero pilot tone
affected by $\varepsilon$ as follows,
\begin{eqnarray}\label{Rk}
\hspace*{-22pt}[\tilde{\mathbf{r}}]_{c_k} &=& \tilde {p}_{c_k}
[\tilde{\mathbf{h}}_N ]_{c_k} \frac{{\sin (\pi \varepsilon )}}
{N\sin(\pi \varepsilon /N)}e^{j\frac{\pi \varepsilon (2 N_g + N - 1)}{N}}\nonumber  \\
& & \!+\! \sum\limits_{k' = 0, k' \ne k }^{N_C - 1}\left\{ \tilde
{p}_{c_{k'}} [\tilde {\mathbf{h}}_N]_{c_{k'}} \frac{{\sin (\pi
\varepsilon )}} {{N\sin [\pi (c_{k'} - c_k + \varepsilon )/N]}}
\right. \nonumber \\
& & \left. \!\times  e^{ j\frac{- \pi (c_{k'} - c_k) + \pi
\varepsilon (2 N_g +N - 1)}{N}} \right\} +
[\tilde{\mathbf{w}}]_{c_k} ,
\end{eqnarray}
where
\begin{equation*}
\tilde{\mathbf{w}} = \mathbf{F}_N^H \mathbf{w}.
\end{equation*}
From (\ref{Rk}), we can see that ICI indeed exists between the two
types of pilot tones. In order to obtain the optimum value of
$\upsilon$, we express $P_{ICI,d_k}$, i.e., the average power of
the total ICI that $\tilde{p}_{d_k }$ imposes on the $N_U$
uniformly spaced pilot tones, as a function of $\upsilon$ as
follows,
\begin{multline}\label{PICI1}
\hspace*{-20pt}P_{ICI,d_k}(\upsilon) \\
= N^{ - 2}E[ |[\mathbf{\tilde{h}}_N ]_{d_k
}| ^2 ] |\tilde{p}_{d_k }|^2  \sin ^2 (\pi \varepsilon )
\sum\limits_{m = 0}^{N_U /2 - 1} {\Upsilon _m (\upsilon) },
\end{multline}
where
\begin{eqnarray}\label{upsilon}
\Upsilon _m (\upsilon) &=& \sin ^{ - 2} [\pi (\upsilon + mX +
\varepsilon )/N] \nonumber \\
& & \!+ \sin ^{ - 2} [\pi (\upsilon - X - mX + \varepsilon )/N].
\end{eqnarray}

    For brevity, we define $\theta _0 = \pi (\upsilon+ mX
+\varepsilon)/N$, $\theta _1 = \pi (\upsilon- X- mX
+\varepsilon)/N$. Based on the definition of $\Upsilon
_m(\upsilon)$, we can obtain the following result,
\begin{multline}
\hspace*{-20pt} \frac{\partial \Upsilon _m(\upsilon)}{\partial
\upsilon} = -\frac{2\pi (\cos \theta_0 \sin ^3 \theta_1 + \cos
\theta_1 \sin
^3 \theta_0) } {N \sin ^3 \theta_0 \sin ^3 \theta_1}  \\
= -{[\sin \theta_0 (\sin \theta_0 - \sin \theta_1 \cos \theta_0
\cos \theta_1) + \cos ^2 \theta_0 \sin ^2 \theta_1]}\\
\times \frac{2\pi \sin (\theta_0 + \theta_1)}{N \sin ^3 \theta_0
\sin ^3 \theta_1} \hfill
\end{multline}
With $\varepsilon \in [-0.5, 0.5]$, $m \in [0, N_U/2-1]$,
$\upsilon \in [1,X-1]$ and $N = X N_U$, we have
\begin{equation*}
\begin{gathered}
\theta_0 \in [\frac{\pi}{2N}, \frac{\pi}{2}-\frac{\pi}{2N} ], \\
\theta_1 \in [-\frac{\pi}{2} +\frac{\pi}{2N}, -\frac{\pi}{2N} ],
\\
\theta_0 +\theta_1 \in [\frac{\pi}{N} -\frac{\pi}{N_U},
\frac{\pi}{N_U}-\frac{\pi}{N}].
\end{gathered}
\end{equation*}
Then the following inequality holds,
\begin{equation}
\sin \theta_0 (\sin \theta_0 - \sin \theta_1 \cos \theta_0 \cos
\theta_1) + \cos ^2 \theta_0 \sin ^2 \theta_1 >0
\end{equation}
Let $\frac{\partial \Upsilon _m(\upsilon)}{\partial \upsilon} = 0
$, we have $\sin (\theta_0 + \theta_1) =0 $, which leads to
$\upsilon = X/2 -\varepsilon$. Besides, we also have
\begin{multline}
\hspace*{-20pt} \frac{\partial ^2 \Upsilon _m(\upsilon)}{\partial
\upsilon ^2} = \frac{2 \pi ^2}{N^2 \sin ^4 \theta_0 \sin ^4 \theta_1 }\\
\times {(\sin ^4 \theta_0 +\sin ^4 \theta_1 +2 \cos ^2 \theta_0
\sin ^4
\theta_1 +2 \cos ^2 \theta_1 \sin ^4 \theta_0)} \\
>0 \hfill
\end{multline}

    Furthermore, we can see from (\ref{upsilon}) that $\Upsilon
_m(\upsilon)$ is symmetrical about the line $\upsilon = X/2-
\varepsilon$. Since $\upsilon$ can only be an integer, the value
of $\upsilon$ within the range $[1, X-1]$ that makes $\Upsilon
_m(\upsilon)$ achieves its minimum for any $\varepsilon$ within
the range $[-0.5, 0.5]$ is $ X/2$.

    Correspondingly, the following result can be obtained,
\begin{equation}
\mathop {\arg \min }\limits_{\upsilon  \in [1,X - 1]}
\{P_{ICI,d_k}(\upsilon)\} = X/2
\end{equation}

    This completes the proof.

\section*{acknowledgements}
    This work was supported by China's High-Tech 863-Future
project under grant 2003AA123310 and National Natural Science
Foundation of China under grant 60496311. The authors would like
to thank the anonymous reviewers for their helpful comments and
also wish to thank Dr. Wenjin Wang for his support concerning
turbo codes.

\profile{Yanxiang Jiang}
   {received the B.S. degree in electrical engineering
from Nanjing University, Nanjing, China, in 1999 and the M.S.
degree in radio engineering from Southeast University, Nanjing,
China, in 2003. He is currently working toward the Ph.D. degree in
National Mobile Communication Research Laboratory, Southeast
University, Nanjing, China. His current research interests include
multicarrier transmission and MIMO systems.}

\profile{Xiqi Gao}
    {received the Ph.D. degree in electrical engineering from Southeast
University, Nanjing, China, in 1997. He joined the Department of
Radio Engineering, Southeast University, in April 1992. From
September 1999 to August 2000, he was a visiting scholar at
Massachusetts Institute of Technology, Cambridge, and Boston
University, Boston, MA. His current research interests include
multicarrier transmission for B3G mobile communications,
space-time coding and spatial multiplexing, iterative
detection/decoding, and signal processing for mobile
communication. He is also a leader of the China High-Tech
863-FuTURE Project. He received the first- and second-class prizes
of the Science and Technology Progress Awards of the State
Education Ministry of China in 1998.}

\profile{Xiaohu You}
    {received the M.S. and Ph.D. degrees from Southeast
University, Nanjing, China, in electrical engineering in 1985 and
1988, respectively. Since 1990 he has been working with National
Mobile Communications Research Laboratory at Southeast University,
where he holds the ranks of professor and director. His research
interests include mobile communications, advanced signal
processing, and applications. He has published two books and over
20 IEEE journal papers in related areas. From 1993 to 1997 he was
engaged, as a team leader, in the development of China's first GSM
and CDMA trial systems. He was the Premier Foundation Investigator
of the China National Science Foundation in 1998. From 1999 to
2001 he was on leave from Southeast University, working as the
chief director of China's 3G (C3G) Mobile Communications R$\&$D
Project. He is currently responsible for organizing China's B3G
R$\&$D activities under the umbrella of the National 863 High-Tech
Program and he is also the chairman of the China 863-FuTURE Expert
Committee.}


\begin{thebibliography}{99}
\providecommand{\url}[1]{#1} \csname url@rmstyle\endcsname
\providecommand{\newblock}{\relax}
\providecommand{\bibinfo}[2]{#2}
\providecommand\BIBentrySTDinterwordspacing{\spaceskip=0pt\relax}
\providecommand\BIBentryALTinterwordstretchfactor{4}
\providecommand\BIBentryALTinterwordspacing{\spaceskip=\fontdimen2\font
plus \BIBentryALTinterwordstretchfactor\fontdimen3\font minus
  \fontdimen4\font\relax}
\providecommand\BIBforeignlanguage[2]{{%
\expandafter\ifx\csname l@#1\endcsname\relax
\typeout{** WARNING: IEEEtran.bst: No hyphenation pattern has been}%
\typeout{** loaded for the language `#1'. Using the pattern for}%
\typeout{** the default language instead.}%
\else \language=\csname l@#1\endcsname \fi #2}}


\bibitem{Bingham}
J.~A.~C. Bingham, ``Multicarrier modulation for data transmission:
{An} idea
  whose time has come,'' \emph{IEEE Commun. Mag.}, vol.~28, pp. 5--14, May
  1990.

\bibitem{Pollet}
T.~Pollet, M.~V. Bladel, and M.~Moeneclaey, ``{BER} sensitivity of
{OFDM}
  systems to carrier frequency offset and {Wiener} phase noise,'' \emph{IEEE
  Trans. Commun.}, vol.~43, pp. 191--193, Feb./Mar./Apr. 1995.

\bibitem{Beek}
J.~van~de Beek, M.~Sandell, and P.~O. Borjesson, ``{ML} estimation
of time and
  frequency offset in {OFDM} systems,'' \emph{IEEE Trans. Signal Processing},
  vol.~45, pp. 1800--1805, July 1997.

\bibitem{Liu}
H.~Liu and U.~Tureli, ``A high-efficiency carrier estimator for
{OFDM}
  communications,'' \emph{IEEE Commun. Lett.}, vol.~2, pp. 104--106, Apr. 1998.

\bibitem{Tureli1}
U.~Tureli, H.~Liu, and M.~D. Zoltowski, ``{OFDM} blind carrier
offset
  estimation: {ESPRIT},'' \emph{IEEE Trans. Commun.}, vol.~48, pp. 1459--1461,
  Sept. 2000.

\bibitem{Moose}
P.~Moose, ``A technique for orthogonal frequency division
multiplexing
  frequency offset correction,'' \emph{IEEE Trans. Commun.}, vol.~42, pp.
  2908--2914, Oct. 1994.

\bibitem{Schmidl}
T.~Schmidl and D.~C. Cox, ``Robust frequency and timing
synchronization for
  {OFDM},'' \emph{IEEE Trans. Commun.}, vol.~45, pp. 1613--1621, Dec. 1997.

\bibitem{Morelli1}
M.~Morelli and U.~Mengali, ``An improved frequency offset
estimator for {OFDM} applications,'' \emph{IEEE Commun. Lett.},
vol.~3, pp. 75--77, Mar. 1999.

\bibitem{Morelli2}
------, ``Carrier-frequency estimation for transmissions over selective
  channels,'' \emph{IEEE Trans. Commun.}, vol.~48, pp. 1580--1589, Sept. 2000.

\bibitem{Xiaoli}
X.~Ma, C.~Tepedelenlioglu, G.~B. Giannakis, and S.~Barbarossa,
``Non-data-aided
  carrier offset estimators for {OFDM} with null sub-carriers: Identifiability,
  algorithms, and performance,'' \emph{IEEE J. Select. Areas Commun.}, vol.~19,
  pp. 2504--2515, Dec. 2001.

\bibitem{Nogami}
H.~Nogami and T.~Nagashima, ``A frequency and timing period
acquisition
  technique for {OFDM} systems,'' in \emph{Proc. PIMRC'95}, vol.~3, Sept. 1995,
  pp. 1010--1015.

\bibitem{Zhang}
R.~Zhang, T.~T. Tjhung, H.~J. Hu, and P.~He, ``Window function and
  interpolation algorithm for {OFDM} frequency-offset correction,'' \emph{IEEE
  Trans. Veh. Technol.}, vol.~52, pp. 654--670, May 2003.

\bibitem{Lei}
J.~Lei and T.-S. Ng, ``A consistent {OFDM} carrier frequency
offset estimator
  based on distinctively spaced pilot tones,'' \emph{IEEE Trans. Wireless
  Commun.}, vol.~3, pp. 588--599, Mar. 2004.

\bibitem{Yanxiang1}
Y.~X. Jiang, D.~M. Wang, X.~Q. Gao, and X.~H. You, ``Low
complexity frequency
  offset estimator for {OFDM} with time-frequency training sequence,'' in
  \emph{Proc. ICC'05}, vol.~4, May 2005, pp. 2553--2557.

\bibitem{Chu}
D.~Chu, ``Polyphase codes with good periodic correlation
properties,''
  \emph{IEEE Tran. Inform. Theory}, vol.~18, pp. 531--532, July 1972.

\bibitem{Negi}
R.~Negi and J.~Cioffi, ``Pilot tone selection for channel
estimation in a mobile {OFDM} system,'' \emph{IEEE Trans. Consumer
Electron.}, vol.~44, pp. 1122--1128,
  Aug. 1998.

\bibitem{Milewski}
A.~Milewski, ``Periodic sequences with optimal properties for
channel
  estimation and fast start-up equalization,'' \emph{IBM J. Res. Develop.},
  vol.~27, pp. 426--431, Sept. 1983.

\bibitem{Miller}
S.~L.~Miller and R.~J. O'Dea, ``Peak power and bandwidth efficient
linear modulation,'' \emph{IEEE Trans. Commun.}, vol.~46, pp.
1639--1648, Dec. 1998.


\bibitem{Tellambura}
C.~Tellambura, ``Computation of the continuous-time {PAR} of an
{OFDM} signal
  with {BPSK} subcarriers,'' \emph{IEEE Commun. Lett.}, vol.~5, pp. 185--187,
  May 2001.

\bibitem{Kay}
S.~M. Kay, \emph{Fundamentals of Statistical Signal Processing:
Estimation
  Theory}, {Englewood Cliffs}~ed.\hskip 1em plus 0.5em minus 0.4em\relax NJ:
  Prentical-Hall, 1993.

\bibitem{Fan}
P.~Fan and W.~H. Mow, ``On optimal training sequence design for
  multiple-antenna systems over dispersive fading channels and its
  extensions,'' \emph{IEEE Trans. Veh. Technol.}, vol.~53, pp. 1623--1626,
  Sept. 2004.

\bibitem{Dolinar}
S. Dolinar and D. Divsalar, ``Weight distributions for turbo codes
using random and nonrandom permutations,'' \emph{NASA JPL TDA
Progress Report}, Tech. Rep.~42, pp. 56--65, Aug. 1995.


\end{thebibliography}
\end{document}